\RequirePackage[displaymath, mathlines]{lineno} 
\documentclass[aps,prl,twocolumn,superscriptaddress,amsmath,amssymb]{revtex4}
\usepackage{graphicx} 
\usepackage{epstopdf} 
\usepackage{dcolumn} 
\usepackage{xcolor} 
\usepackage{amsmath,amssymb} 
\usepackage{hyperref} 
\usepackage[utf8]{inputenc} 
\usepackage{blindtext} 
\usepackage{ifthen} 
\usepackage{orcidlink} 

\graphicspath{{figures/}} 

\newboolean{articletitles}
\setboolean{articletitles}{true} 

\input{belle2-symbols}

\ifthenelse{\isundefined{\analysis}}{
    
}{

}

\newboolean{wordcount}
\setboolean{wordcount}{false} 

\ifthenelse{\boolean{wordcount}}%
{\usepackage{bibentry} 
 \usepackage{comment} 
 \excludecomment{align*} 
 \excludecomment{align} 
 \excludecomment{equation*} 
 \excludecomment{eqnarray*} 
 \excludecomment\textit{}{eqnarray} 
 \excludecomment{acknowledgments} 
 \def\maketitle{} 
}{}

\begin{document}
\title{
First search for $B \rightarrow X_{s} \nu \bar{\nu}$ decays
}


\ifthenelse{\boolean{wordcount}}{}{
  \author{M.~Abumusabh\,\orcidlink{0009-0004-1031-5425}} 
  \author{I.~Adachi\,\orcidlink{0000-0003-2287-0173}} 
  \author{K.~Adamczyk\,\orcidlink{0000-0001-6208-0876}} 
  \author{L.~Aggarwal\,\orcidlink{0000-0002-0909-7537}} 
  \author{H.~Ahmed\,\orcidlink{0000-0003-3976-7498}} 
  \author{Y.~Ahn\,\orcidlink{0000-0001-6820-0576}} 
  \author{H.~Aihara\,\orcidlink{0000-0002-1907-5964}} 
  \author{N.~Akopov\,\orcidlink{0000-0002-4425-2096}} 
  \author{S.~Alghamdi\,\orcidlink{0000-0001-7609-112X}} 
  \author{M.~Alhakami\,\orcidlink{0000-0002-2234-8628}} 
% ...
\collaboration{The Belle II Collaboration}

}

\begin{abstract}
We report the first search for the flavor-changing neutral-current decays $B \rightarrow X_{s} \nu \bar{\nu}$, where $X_{s}$ is a hadronic system with strangeness equal to 1, in data collected with the Belle~II detector at the SuperKEKB asymmetric-energy $e^+e^-$ collider.
The data sample corresponds to an integrated luminosity of $365\invfb$ collected at the $\FourS$ resonance and $43\invfb$ collected at a center-of-mass energy $60\mev$ below resonance for estimation of $e^+e^-\to q\bar{q}$ continuum background.
One of the $B$ mesons from the $\FourS \to B\Bbar$ decay is fully reconstructed in a hadronic decay mode. 
The $B \to X_s \nu \bar{\nu}$ decay is reconstructed with a sum-of-exclusives approach that uses 30 $X_s$ decay modes. This approach provides high sensitivity to the inclusive decay, despite the presence of two undetected neutrinos.
The search is performed in three regions of the $X_{s}$ mass, $M_{X_{s}}^{\textrm{true}}$, chosen to separate contributions from $K$, $K^{*}(892)$, and heavier strange final states.
We do not observe a significant signal and set upper limits at 90\% confidence level on the partial branching fractions for the regions $0.0 < M_{X_{s}}^{\textrm{true}} < 0.6\gevcc$, $0.6 < M_{X_{s}}^{\textrm{true}} < 1.0\gevcc$, and $1.0\gevcc < M_{X_{s}}^{\textrm{true}}$ of $2.2 \times 10^{-5}$, $9.3 \times 10^{-5}$, and $30.9 \times 10^{-5}$, respectively. 
Combining the three mass regions, we obtain the upper limit on the branching fraction, $B(B \to X_s \nu \bar{\nu}) < 3.3 \times 10^{-4}$.
\end{abstract}

\maketitle

\begin{figure}[htb]
  \centering
  \begin{tabular}{c @{\qquad} c }
    \includegraphics[width=0.22\textwidth]{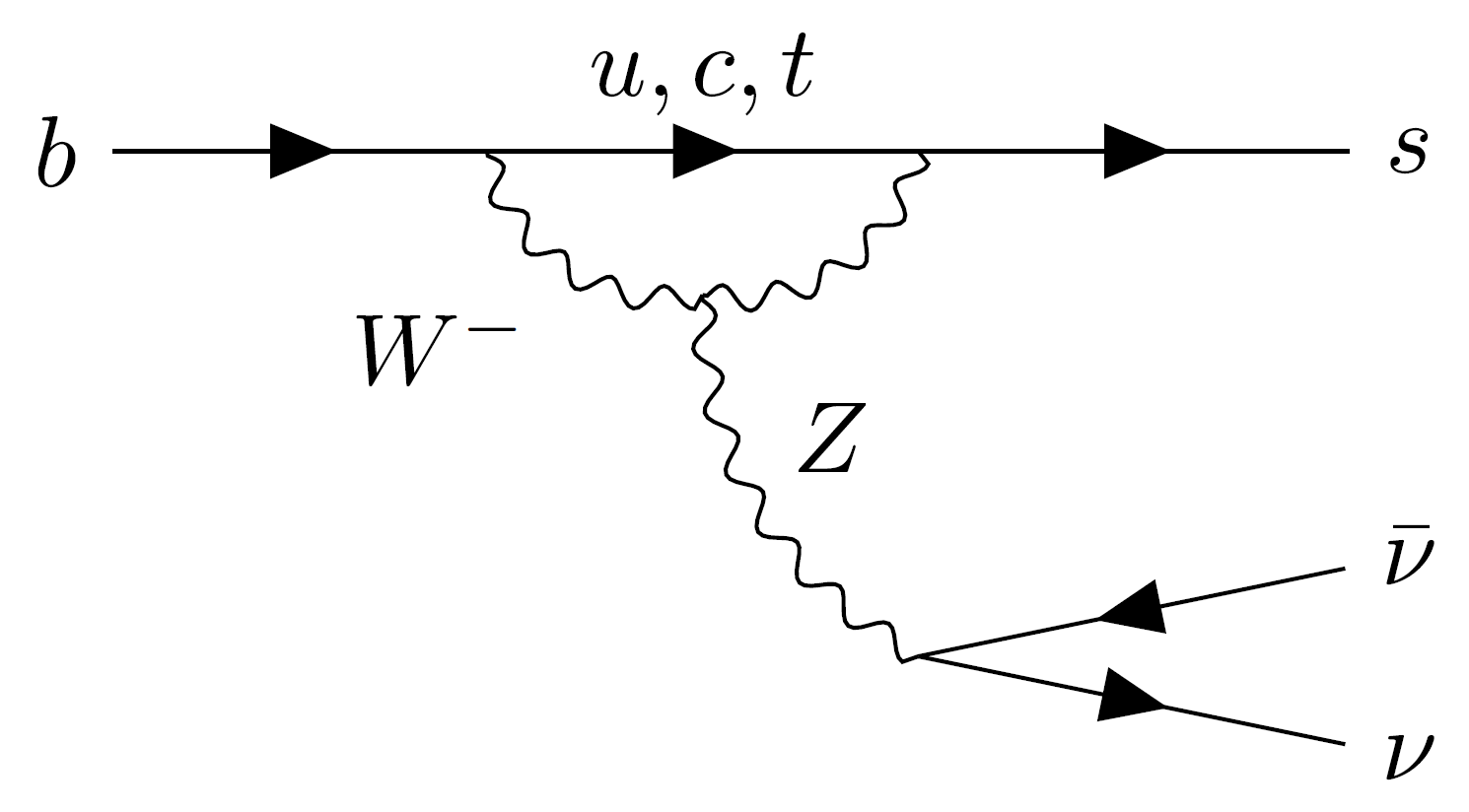} &
    \includegraphics[width=0.22\textwidth]{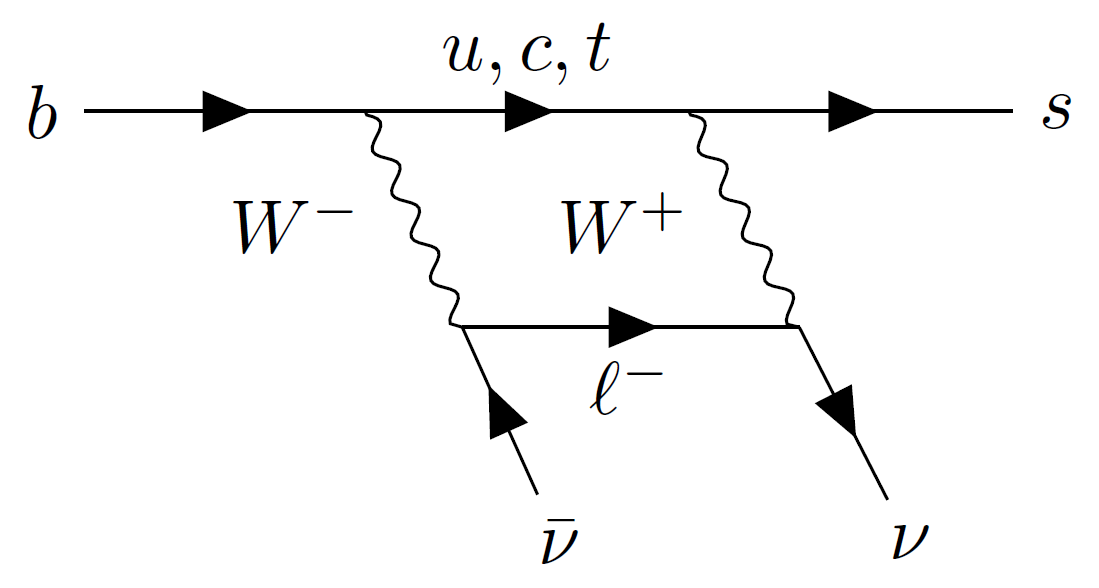} \\
    \small (a) Penguin diagram & \small (b) Box diagram
  \end{tabular}
    \caption{\label{fig:diagram} The leading order SM diagrams for quark-level $b\rightarrow s \nu \bar{\nu}$ decays.}
\end{figure}

Flavor changing neutral current $b\rightarrow s \nu \bar{\nu}$ decays are suppressed in the Standard Model~(SM) due to the GIM mechanism~\cite{Glashow:1970gm}, and occur only via loop diagrams~(Fig.~\ref{fig:diagram}).
The absence of charged leptons in the final state allows precise theoretical predictions, in contrast to $b \to s \ell^+ \ell^-$ decays, which have contributions from photon exchange between charged particles and a charm loop~\cite{Khodjamirian:2010vf}.
The decay rates can be enhanced by physics beyond the SM; for example, virtual contributions from heavy particles such as leptoquarks~\cite{PhysRevD.95.035027} or a $Z^{'}$ boson~\cite{Buras:2012jb}.
The experimental signature is also sensitive to new invisible particles, such as invisible light scalars~\cite{Calcuttawala:2017usw} or fermionic dark matter~\cite{Buras:2024ewl}.

The Belle~II collaboration found evidence for the exclusive decay $B^{+} \rightarrow K^{+} \nu \bar{\nu}$, measuring a branching fraction ${\cal B}(B^{+}\to K^{+}\nu\bar{\nu})=(2.3 \pm 0.7 ) \times 10^{-5}$~\cite{Belle-II:2023esi}, which is 2.7 standard deviations above the SM prediction, $(5.04 \pm 0.40) \times 10^{-6} $~\cite{Parrott:2022zte,Belle-II:2025lfq}.
Inclusive $B \to X_s \nu \bar{\nu}$ decays, where $X_s$ denotes a hadronic system with strangeness equal to 1, have not been studied.
The branching fraction ${\cal{B}}(B\rightarrow X_{s} \nu \bar{\nu})$ is predicted to be $\left(2.9 \pm 0.3 \right) \times 10^{-5}$ in the SM~\cite{Buras:2014fpa}, excluding the long-distance contribution from $B^{+} \rightarrow \tau^{+}(\rightarrow X_{s}^{+} \bar{\nu}) \nu$ decay.
The inclusive $B \to X_s \nu \bar{\nu}$ and exclusive $B \rightarrow K \nu \bar{\nu}$ branching fractions have different dependence on the parameters of physics beyond the SM and can be related with a sum rule~\cite{Felkl:2021uxi,Buras:2024ewl}, which provides strong motivation for the inclusive search.
The ALEPH collaboration searched for $b$-hadrons decaying with large missing energy at the $Z$-pole and set an upper limit on ${\cal{B}}(b \to s \nu \bar{\nu})$ of $6.4\times 10^{-4}$ at 90\% confidence level~(C.L.)~\cite{ALEPH:2000vvi}. However, unlike for $B$ mesons, there is no established theoretical framework that provides reliable predictions for inclusive $b$-hadron decays.

In this Letter, we report the first search for $B \rightarrow X_{s} \nu \bar{\nu}$, where $B$ and $X_{s}$ denote both charged and neutral states.
The data sample consists of $365.4\pm1.7\invfb$ and contains $(387.1\pm5.6)\times10^{6}$ $\FourS \rightarrow B\Bbar$ decays collected by the Belle II~\cite{Abe:2010gxa} detector at the SuperKEKB asymmetric-energy $e^+e^-$ collider~\cite{Akai:2018mbz}.
An additional sample of $42.7\pm0.2$\invfb collected at an energy 60\mev below the $\FourS$ peak (off-resonance data) is used for assessing the background from the continuum $e^{+}e^{-}\to q\bar{q}$ process, where $q$ is a $u$, $d$, $s$, or $c$ quark.

The Belle~II detector consists of a two-layer silicon pixel detector~(PXD), a four-layer silicon vertex detector, a central drift chamber~(CDC),
a time of propagation~(TOP) counter, an aerogel ring-imaging Cherenkov detector~(ARICH), an electromagnetic calorimeter~(ECL), a superconducting solenoid, and a $\KL$ and muon detector~(KLM). The symmetry axis of these detectors, defined as the $z$-axis, is approximately in the direction of the electron beam~\cite{Abe:2010gxa}.

The Belle II software framework~(\textsc{basf2})~\cite{Kuhr:2018lps, basf2-zenodo} is used for event reconstruction.
Monte Carlo simulated samples are used to optimize the analysis.
$B$-meson production and decay are simulated using the \textsc{EvtGen}~\cite{Lange:2001uf} program; $q\bar{q}$ background samples are generated with the \textsc{Pythia8}~\cite{Sjostrand:2014zea} and \textsc{KKMC}~\cite{Jadach:1999vf} software; photon radiation from charged particles is simulated using the \textsc{PHOTOS}~\cite{Barberio:1990ms} package; and the \textsc{Geant4} program is used to simulate the detector response to the passage of particles~\cite{Agostinelli:2002hh}.

We produce signal simulation samples consisting of individually generated contributions: exclusive $B \rightarrow K \nu \bar{\nu}$ and $B \rightarrow K^{*} \nu \bar{\nu}$ decays, where $K^{*}$ denotes the $K^{*}(892)$ resonance, and non-resonant $B \rightarrow X_{s} \nu \bar{\nu}$ decays.
The $B \rightarrow K \nu \bar{\nu}$ and $B \rightarrow K^{*} \nu \bar{\nu}$ decays are generated using form factors calculated in lattice QCD computations~\cite{Parrott:2022zte, Parrott:2022rgu} or light-cone sum rules~\cite{Bharucha:2015bzk}.
The non-resonant $B \rightarrow X_{s} \nu \bar{\nu}$ sample is generated with the requirement that the mass $M_{X_{s}}^{\textrm{true}}$ of the $X_{s}$ system be greater than $1.1~\gevcc$.
This requirement is motivated by the study of the $B \rightarrow X_{s} \gamma$ decay~\cite{PhysRevD.91.052004}, as well as $B^{0} \rightarrow K^{+}\pi^{-}\mu^{+}\mu^{-}$ and $B^{+} \rightarrow K^{+}\pi^{+}\pi^{-}\mu^{+}\mu^{-}$ decays, which do not show clear components below this mass threshold except $K^{*}$ resonance~\cite{LHCb:2016ykl,LHCb:2014osj}.
The exclusive $K^{*}\nu\bar{\nu}$ samples are generated without restrictions on the mass $M_{X_{s}}^{\textrm{true}}$.
To simulate the $M_{X_{s}}^{\textrm{true}}$ distribution, we use the Fermi motion model~\cite{Ali:1996bm,Kim:1998wx,HFLAV:2022pwe}, assuming the momentum of the $b$-quark follows a Gaussian distribution.
The fragmentation of the final-state hadrons is simulated using \textsc{Pythia8}.
The invariant mass squared of the neutrino pair, $q^{2}$, is simulated according to the SM~\cite{Altmannshofer:2009ma}.
The long-distance $B^{+} \rightarrow \tau^{+}(\rightarrow X_{s}^{+} \bar{\nu}) \nu$ decays are treated as background and included in the background sample.

One $B$ meson, denoted $B_{\rm tag}$, is fully reconstructed in a hadronic decay mode by the full event interpretation~(FEI) algorithm~\cite{Keck:2018lcd}.
The FEI assigns a probability to each $B_{\rm tag}$ candidate based on a multivariate classifier, indicating how likely the candidate is to be properly reconstructed in a given mode.
Candidates with a probability greater than 0.001 are selected.
We require $B_{\rm tag}$ candidates to satisfy selections on the beam-energy-constrained mass, $M_{\textrm{bc}} \equiv \sqrt{\left( \sqrt{s}/2 \right)^{2} - \left| p_{B}^{*} c\right|^{2}}/c^{2} > 5.27\gevcc$, and the energy difference, $\left| \Delta E \right| \equiv \left| E_{B}^{*} - (\sqrt{s}/2) \right| < 0.2\gev$.
Here, $\sqrt{s}$, $p_{B}^{*}$, and $E_{B}^{*}$ are the SuperKEKB center-of-mass~(c.m.) energy, and the momentum and energy of the $B$ meson in the c.m. system, respectively.

Particle candidates not used to reconstruct the $B_{\textrm{tag}}$ are used to form the $X_{s}$ candidate.
We use a sum-of-exclusive decays method to reconstruct the $X_{s}$ candidate, which provides high sensitivity to the inclusive decay, despite the presence of two undetected neutrinos.
The $X_{s}$ candidate is reconstructed in 30 decay modes: $Kn\pi$ $(0 \leq n \leq 4)$ with at most two $\pi^{0}$, and $3Kn\pi$ $(0 \leq n \leq 1)$ with at most one $\pi^0$ and one $\KS$, where $K$ indicates $K^{\pm}$ or $\KS$.
These decay modes cover 83\% of the non-resonant $X_{s}$ decay in the signal simulation samples, assuming the $\KL$ contribution is equal to that of $\KS$.
In total, the reconstructed modes constitute 93\% of the $B \to X_{s} \nu \bar{\nu}$ branching fraction in the signal simulation samples.

The $X_{s}$-candidate reconstruction proceeds as follows.
Except for $\pi^{\pm}$ from $\KS$ candidates, the distance of closest approach of a track to the interaction point~(IP) is required to satisfy $dr < 0.5\cm$ in the transverse plane and $\left| dz \right| < 2\cm$ along the $z$-axis.
The numbers of hits in the PXD and CDC must be greater than 0 and 20, respectively.
Kaon and pion candidates are selected using charged particle identification~(PID) likelihoods, built using information from the SVD, CDC, TOP, ARICH, ECL, and KLM subsystems.

Photon candidates are reconstructed from ECL clusters not geometrically matched to charged tracks; to guarantee that clusters do not originate from charged particles, their polar angle $\theta$ must be within the CDC acceptance, $17\degrees < \theta < 150\degrees$.
The energy of a photon candidate must be greater than 80\mev, 30\mev, or 60\mev for candidates detected in the forward, barrel, and backward regions of the ECL, respectively.

We reconstruct $\pi^{0}$ candidates from pairs of photon candidates with invariant mass in the range $120 < m_{\gamma\gamma} < 145~\mevcc$.
The momentum of $\pi^{0}$ candidates must be larger than $0.4$$\gevc$.
We require the absolute difference in azimuthal angle between the two photons to be less than $1.5$~$\textrm{rad}$ in the laboratory frame. Additionally, the opening angle must be smaller than $1.4~\textrm{rad}$.

We reconstruct $\KS$ candidates from two oppositely charged tracks, for which a vertex fit~\cite{Krohn:2019dlq} is performed and required to converge.
The significance of the distance from this vertex to the IP must be larger than 50.
We require the invariant mass of $\KS$-candidates to be within $10\mevcc$ of the nominal $\KS$ mass~\cite{ParticleDataGroup:2024cfk}.

To suppress combinatorial background, the reconstructed invariant mass of the $X_{s}$ candidate~($M_{X_{s}}^{\rm reco}$) must be lower than $2.0\gevcc$.
A decay-chain vertex fit~\cite{Krohn:2019dlq} is required to converge for $X_{s}$ candidates in decay modes with two or more charged tracks.
The c.m.-frame momentum of the $X_{s}$ must satisfy $0.5 < p^{*}_{X_{s}} < 2.96\gevc$.
To suppress backgrounds from $D$ mesons, $X_{s}$ candidates with $M_{X_{s}}^{\textrm{reco}}$ in the range $[1.84, 1.89]$$\gevcc$ are rejected.  The $X_s^0$ flavor is assigned based on the net charge of the associated kaons; if zero, the flavor is undetermined.
Pairs of $X_s$ and $B_{\textrm{tag}}$ candidates are further considered only if they have total charge zero and opposite flavor (where determined).

The dineutrino system can be mimicked by particles outside the detector acceptance.
Therefore, we require the polar angle of the missing momentum $\vec{p}_{\textrm{miss}} = \vec{p}_{e^{+}e^{-}} - \sum\vec{p}_{\textrm{track}} - \sum\vec{p}_{\gamma}$ to be in the detector acceptance, $17\degrees < \theta_{\textrm{miss}} < 150\degrees$. Here, $\vec{p}_{\textrm{track}}$ and $\vec{p}_{\gamma}$ are the momenta of charged tracks and photon candidates.
 
For each combination of $X_s$ and $B_{\textrm{tag}}$ candidates, the tracks or photon candidates not used in the reconstruction of these candidates form the rest-of-event.
The numbers of tracks from the IP, $\pi^{0}$, and $\KS$ candidates in the rest-of-event are required to be zero.
The rest-of-event is allowed to contain photon candidates and tracks with $dr > 2~\cm$ or $\left| dz \right| > 4~\cm$.
We calculate the total lab-frame energy $E_{\textrm{extra}}$ as the sum of all photon candidates with energies greater than 100, 60, and 150\mev in the forward, barrel, and backward regions of the ECL, respectively. $E_{\textrm{extra}}$ is required to be less than $1.3\gev$, and is further used for background suppression, as explained later.

After these selections, the average multiplicity of $B_{\rm tag}X_{s}$ candidates per event is 1.71 in simulated signal samples and 1.45 in data.
In each event, we retain only the candidates with the highest $B_{\textrm{tag}}$ FEI probability.
After this, the average multiplicity is reduced to 1.06 for simulated signal samples and 1.13 for data, with negligible statistical uncertainty.
Multiple candidates arise from the exchange of photons or tracks that are displaced from the IP, which are not vetoed in the rest-of-event.
If multiple signal candidates remain, we select one at random.

The dominant background contributions arise from continuum $e^{+}e^{-} \to q\bar{q}$ processes and non-signal $B\Bbar$ decays.
To suppress them, we use a boosted decision tree (BDT)~\cite{Keck:2017gsv}.
The BDT uses 32 input variables: 26 variables 
related to kinematics, the rest-of-event, and event shape, selected for their separation power and good data-simulation agreement, and 6 variables specifically designed to suppress backgrounds from $D$ mesons.
The list of input variables is provided in the Supplemental Material~\cite{supp}.

Among the 32 BDT-input variables, $E_{\textrm{extra}}$ provides the largest signal-background discrimination.
We require the BDT output, denoted $\mathcal{O}_{\textrm{BDT}}$ and ranging from 0 to 1, to be greater than 0.86.
This criterion is chosen to include the point that maximizes the figure of merit~(FOM), defined as $N_{\textrm{sig}}/\sqrt{N_{\textrm{sig}}+N_{\textrm{bkg}}}$ where $N_{\textrm{sig}}$ and $N_{\textrm{bkg}}$ are the expected numbers of signal and background events.
According to the simulation, the $\mathcal{O}_{\textrm{BDT}}$ requirement retains 68\% of signal events while rejecting 97\% of background.
The expected background yield is 3338 events, comprising 1568 $q\bar{q}$ and 1770 $B\Bbar$ events. This total includes $4.0\pm0.9$ events from the $B^{+} \rightarrow \tau^{+}(\rightarrow X_{s}^{+}\bar{\nu})\nu$ background.
In the simulated signal sample, the ratio of charged to neutral $B$ decay modes is approximately 2.7:1.

A signal region is defined in $M_{X_{s}}^{\textrm{reco}} \times {\mathcal{O}_{\textrm{BDT}}}$ space and divided into $3 \times 5$~bins.
The bin boundaries along the $M_{X_{s}}^{\textrm{reco}}$ axis, $[0.0, 0.6, 1.0, 2.0]\gevcc$, correspond to regions dominated by $K$, $K^{*}$, and higher-mass $X_{s}$ states, labeled MR1, MR2, and MR3.
The boundary between MR2 and MR3 is set to $1.0\gevcc$, slightly below the non-resonant generation threshold of $1.1\gevcc$, to minimize sensitivity to the modeling of the non-resonant $X_{s}$ generation point.
In MR1, MR2, and MR3, the bin boundaries along the ${\mathcal{O}_{\textrm{BDT}}}$ axis are [0.86, 0.9195, 0.979, 0.986, 0.993, 1.0], [0.86, 0.8925, 0.925, 0.95, 0.975, 1.0], and [0.86, 0.9, 0.94, 0.96, 0.98, 1.0], respectively.
These bin boundaries are chosen to maximize sensitivity without introducing fit bias, which is verified by a pseudo-experiment study.
For simplicity, bin indices are assigned from 1 to 15.
The indices range from 1 to 5 in MR1, from 6 to 10 in MR2, and from 11 to 15 in MR3, increasing along the positive $\mathcal{O}_{\textrm{BDT}}$ axis within each mass region.

We perform a binned maximum likelihood fit to the $M_{X_{s}}^{\textrm{reco}} \times {\mathcal{O}_{\textrm{BDT}}}$ distribution to determine the signal yields in three $M_{X_{s}}^{\textrm{true}}$ regions; $0.0 < M_{X_{s}}^{\textrm{true}} < 0.6\gevcc$, $0.6 \leq M_{X_{s}}^{\textrm{true}} < 1.0\gevcc$ and $1.0\gevcc \leq M_{X_{s}}^{\textrm{true}}$, where $M_{X_{s}}^{\textrm{true}}$ is the true mass of the $X_{s}$ system, obtained from the simulation.
The relationship between reconstructed and true mass, including cross-feed between regions, is derived from the signal simulation.
We use the \textsc{HistFactory}~\cite{Cranmer:2012} package to construct the probability density function~(PDF) as the product of Poisson distributions and auxiliary terms.
The expected numbers of events for the Poisson distributions are determined from simulation and includes correction factors and systematic uncertainties, described below.
We introduce nuisance parameters to the likelihood with Gaussian constraints and parametrize systematic uncertainties as auxiliary terms.
The values of the nuisance parameters and the signal branching fractions are determined by the fit.

Data-driven corrections are applied to the modeling.
The $B^{+} \rightarrow K^{+} \nu \bar{\nu}$ analysis~\cite{Belle-II:2023esi} observed a mismodeling of the photon multiplicity in Belle~II.
This mismodeling likely arises from beam-related backgrounds and energy deposits from hadrons.
To address this, we compare the photon multiplicity distributions in data and simulation using $B_{\textrm{tag}}^{\pm}X_{s}^{0}$ candidates formed by combining a charged $B_{\textrm{tag}}$ candidate and a neutral $X_{s}$ candidate.
Correction factors are derived from the ratio of data to simulation.
We reweight simulation samples as a function of the photon multiplicity to correct for the disagreement.
The simulation of the fragmentation of the $X_{s}$ system is corrected using results from a measurement of $B \rightarrow X_{s} \gamma$~\cite{PhysRevD.99.032012} in three $M_{X_{s}}^{\textrm{true}}$ regions: $[1.15, 1.5]\gevcc$, $[1.5, 2.0]\gevcc$, and $[2.0, 2.4]\gevcc$.
Corrections are applied to nine decay mode categories: $K\pi$ without $\pi^{0}$, $K\pi$ with one $\pi^{0}$, $K2\pi$ without $\pi^{0}$, $K2\pi$ with one $\pi^{0}$, $K3\pi$ without $\pi^{0}$, $K3\pi$ with one $\pi^{0}$, $K4\pi$ with at most one $\pi^{0}$, decay modes including $2\pi^{0}$, and decay modes including $3K$.
The $B \rightarrow X_{s} n\bar{n}$ and $B \rightarrow X_{s} \KL \KL$ decays constitute peaking backgrounds and require accurate modeling.
For the modeling of $B \rightarrow K^{(*)} n\bar{n}$ decays, branching fractions are corrected using measurements of $B \rightarrow K^{(*)} p \bar{p}$ decays~\cite{BaBar:2007esv,Belle:2008zkc,Belle:2007oni}, and the distribution of neutron-pair invariant mass, $M_{n\bar{n}}$, is modeled using the distribution of proton-pair mass, $M_{p\bar{p}}$, from these decays.
To correct the remaining $B \rightarrow X_{s} n\bar{n}$ simulation samples, we reconstruct $B \rightarrow X_{s} p\bar{p}$ events in data and use the measured $M_{p\bar{p}}$ distribution to reweight the simulated $M_{n\bar{n}}$ shape.
The 
kinematics
and branching fractions of $B^{+} \rightarrow K^{+} \KL \KL$ and $B^{0} \rightarrow \KS \KL \KL$ are corrected based on the observed $B^{+} \rightarrow K^{+} \KS \KS$~\cite{BaBar:2012iuj} and $B^{0} \rightarrow \KS \KS \KS$~\cite{BaBar:2011ktx, Gronau:2005ax} distributions, respectively.
The modeling of $B \rightarrow K^{*} \KL \KL$ and the remaining $B \rightarrow X_{s} \KL \KL$ decays is corrected using reconstructed $B \rightarrow K^{*} \KS \KS$ and $B \rightarrow X_{s} \KS \KS$ decays.

Four samples are used for corrections to and validation of the PDFs, and for estimating the PDF-related systematic uncertainties: the off-resonance data, an $M_{\textrm{bc}}$ sideband, an ${\mathcal{O}_{\textrm{BDT}}}$ sideband, and a $B \rightarrow J/\psi X_{s}$ sample.
To estimate the systematic uncertainty for the ${\mathcal{O}_{\textrm{BDT}}}$ distribution for $e^{+}e^{-}\rightarrow q \bar{q}$ background, a BDT with the same input variables is trained to distinguish off-resonance data from simulation~\cite{Martschei_2012}.
A weight $\sqrt{t/(1-t)}$, where $t$ is the BDT output, is applied to each event of the on-resonance $e^{+}e^{-} \rightarrow q \bar{q}$ simulation sample.
The resulting ${\mathcal{O}_{\textrm{BDT}}}$ distribution is used in the fit, and the changes in the signal branching fractions are taken as systematic uncertainties.
Additionally, data-to-simulation efficiency ratios are measured to be $1.21 \pm 0.13$, $0.90 \pm 0.20$, and $0.78 \pm 0.10$ for MR1, MR2, and MR3, respectively.
These ratios are applied as correction factors to the $q\bar{q}$ efficiencies, and their statistical uncertainties are assigned as systematic uncertainties.
The ${\mathcal{O}_{\textrm{BDT}}}$ distribution for $B\Bbar$ events is validated with the $M_{\textrm{bc}}$ sideband, defined as $5.20\gevcc < M_{\textrm{bc}} < 5.26\gevcc$, and with the ${\mathcal{O}_{\textrm{BDT}}}$ sideband, defined as $0.80 < {\mathcal{O}_{\textrm{BDT}}} < 0.86$.
We compare the efficiencies of the $\mathcal{O}_{\textrm{BDT}}$ requirement between data and simulation samples using $B\rightarrow J/\psi X_{s}$ events with $J/\psi \rightarrow \mu^{+} \mu^{-}$.
The reconstruction and selection procedures are the same as for the $B\rightarrow X_{s} \nu \bar{\nu}$ analysis, 
except for the selection of a $J/\psi\to\mu^{+}\mu^{-}$ candidate.
We require the invariant mass of the $J/\psi$ candidate to be within $0.05\gevcc$ of the nominal mass~\cite{ParticleDataGroup:2024cfk}.
The $B$ candidate from the $B\rightarrow J/\psi X_{s}$ decay must satisfy $M_{\rm bc} > 5.25\gevcc$ and $\left| \Delta E \right| < 0.1\gev$.
The event is then treated as a $B \to X_{s} \nu \bar{\nu}$ event by ignoring the muons from $J/\psi \to \mu^{+} \mu^{-}$.
We verified using simulation that the $\mathcal{O}_{\textrm{BDT}}$ selection efficiencies are consistent between $B\rightarrow J/\psi X_{s}$ and signal decays for MR1 and MR2.
The data-to-simulation $\mathcal{O}_{\textrm{BDT}}$ selection efficiency ratios are $1.00 \pm 0.04$, $1.05 \pm 0.08$, and $0.97 \pm 0.14$ for MR1, MR2, and MR3, respectively.
These ratios are used as correction factors, and their statistical uncertainties are assigned as systematic uncertainties.

\begin{table}[thb]
\renewcommand{\arraystretch}{1.2}
\caption{Summary of the dominant systematic uncertainties and the total contribution from subdominant uncertainties on ${\cal{B}}(B \to X_s \nu \bar{\nu})$ in the entire $M_{X_{s}}^{\textrm{true}}$ range. The impact on the branching fraction uncertainty $\sigma_{\cal{B}}$ from each source is estimated by fixing the corresponding nuisance parameter in the minimization procedure and subtracting its uncertainty in quadrature from the total uncertainty. Due to correlations among sources, the quadrature sum of individual 
uncertainties does not equal the total 
uncertainty.}
\label{tab:uncertaintiestable}
\centering
\begin{tabular*}{\linewidth}{@{\extracolsep{\fill}}lc}
 \hline
  \hline
Source & Impact on $\sigma_{\cal{B}}$ [$10^{-5}$] \\
 \hline
Simulated-sample size & 6.0\\
Background normalization & 5.7\\
Branching fractions of major $B$-decays & 2.3\\
Non-resonant $X_{s}\nu\bar{\nu}$ generation range & 2.1\\
$\mathcal{O}_{\textrm{BDT}}$ selection efficiency & 2.1\\
Photon multiplicity correction & 1.8\\
$q\bar{q}$ background efficiency & 1.8\\
Other subdominant contributions & 3.1\\
\hline
Total systematic sources & 11.7\\
 \hline
 \hline
\end{tabular*}
\end{table}
Table~\ref{tab:uncertaintiestable} shows a summary of the dominant systematic uncertainties.
The finite size of the simulation samples used to determine the PDF shapes is included as a source of systematic uncertainty, and is dominated by the limited statistics of the background simulation samples.
The ratio of data to simulation in the $M_{\textrm{bc}}$ and ${\mathcal{O}_{\textrm{BDT}}}$ sideband regions shows a maximum deviation of 12\%.
A conservative systematic uncertainty of $\pm20\%$ is assigned to the background normalization in each $M_{X_s}^{\textrm{reco}}$ region: MR1, MR2, and MR3.
To estimate the systematic uncertainty of the photon multiplicity correction factor, we compare the photon multiplicity distributions in data and simulation with $B_{\textrm{tag}}^{0}X_{s}^{\pm}$ candidates.
The residual difference is assigned as the systematic uncertainty of photon multiplicity correction factor.
The lower bound on the $M_{X_{s}}^{\textrm{true}}$ mass 
when generating the non-resonant $B \to X_{s} \nu \bar{\nu}$ component is set to $1.1\gevcc$.
We assign a systematic uncertainty of $\pm0.1\gevcc$ to this value, which is conservatively estimated from the $M_{X_{s}}^{\textrm{true}}$ distribution of $B \rightarrow X_{s} \gamma$, $B^{0} \rightarrow K^{+}\pi^{-}\mu^{+}\mu^{-}$, and $B^{+} \rightarrow K^{+}\pi^{+}\pi^{-}\mu^{+}\mu^{-}$ decays~\cite{PhysRevD.91.052004,LHCb:2016ykl,LHCb:2014osj}.
The impact of an S-wave $K\pi$ component in the $M_{X_{s}}^{\textrm{true}} < 1.0\gevcc$ region was determined with simulation to have a negligible effect.
The impact of varying the branching fractions of the leading $B$ meson decays~\cite{ParticleDataGroup:2024cfk} is also included as a systematic uncertainty.
The full list of contributions to the systematic uncertainty, including subdominant contributions, is included in the Supplemental Material~\cite{supp}.

\begin{figure}[h]
\centering
    \includegraphics[width=\columnwidth]{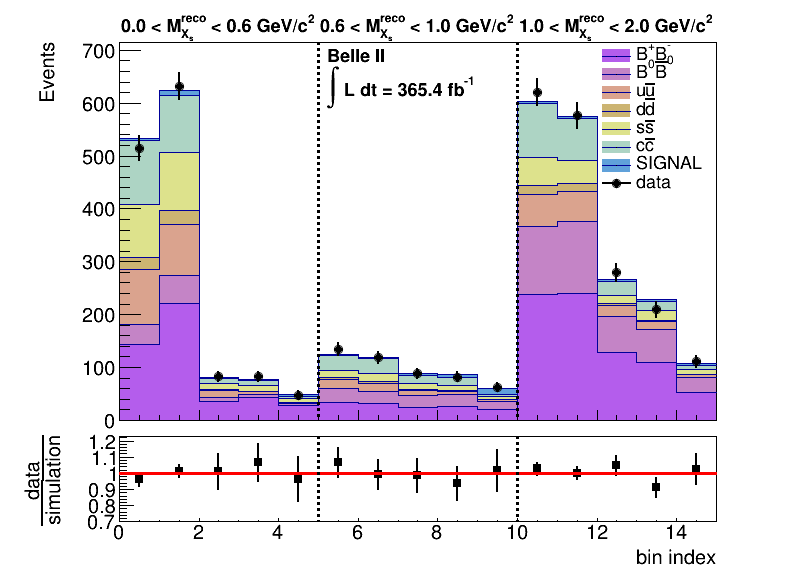}
    \caption{\label{fig:postfit} The bin index distribution after the fit, for data and histogram templates. The background templates are separated into 6~categories: charged $B$ meson decay, neutral $B$ meson decay, and four types of $e^{+}e^{-} \rightarrow q\bar{q}$ background~$\left(q=u,d,s,c \right)$.}
\end{figure}

The post-fit distribution of the bin index is shown in Fig.~\ref{fig:postfit}.
The branching fractions, given in Table~\ref{tab:bf}, are calculated as $\mathcal{B} = N_{\rm sig}/(2 \times N_{BB} \times \epsilon)$, where $N_{BB}$ is determined as in Ref.~\cite{Belle:2025nup}.
The signal distribution is shown separately in the Supplemental Material~\cite{supp}.
The observed signal yields are not significant, and upper limits~(ULs) on the branching fractions are determined using the $\textrm{CL}_{s}$ method~\cite{Read:2002hq}.
The $\textrm{CL}_{s}$ values are shown in the Supplemental Material~\cite{supp}.
The central value of ${\cal B}(B\to K\nu\bar{\nu})$
determined from the fit results
is $[0.3\pm0.8(\rm stat)^{+0.8}_{-0.7}(\rm syst)]\times 10^{-5}$, consistent with the value $[1.1^{+0.9}_{-0.8}(\rm stat)^{+0.8}_{-0.5}(\rm syst)]\times 10^{-5}$ obtained in the dedicated $B^{+} \rightarrow K^{+} \nu \bar{\nu}$ analysis~\cite{Belle-II:2023esi}.

\begin{table}[ht]
\renewcommand{\arraystretch}{1.3}
      \caption{Efficiencies~($\epsilon$), signal yields~($N_{\rm sig}$), and branching-fraction~($\mathcal{B}$) central values and upper limits, where the subscripts obs and exp indicate observed and expected values, respectively. The efficiencies are determined from the fit, since they depend on nuisance parameters. 
    The expected upper limit is calculated from the background-only hypothesis. The first and second uncertainties are statistical and systematic, respectively. }
  \label{tab:bf}
  \centering
  \begin{tabular*}{\linewidth}{@{\extracolsep{\fill}}lccrrr}
\hline \hline
                &            &                &         \multicolumn{3}{c}{$\mathcal{B}$ [$10^{-5}$]} \\\cline{4-6}
$M_{X_{s}}^{\textrm{true}}$$\left[ \textrm{GeV}/c^{2} \right]$ & $\epsilon$$\left[ 10^{-3} \right]$ & $N_{\rm sig}$  & central value & $\textrm{UL}_{\textrm{obs}}$ & $\textrm{UL}_{\textrm{exp}}$ \\
\hline
$[0, 0.6]$                          &  2.93    &  $6^{+18}_{-17}{}^{+19}_{-16}$   &   $0.3\pm0.8 {}^{+0.8}_{-0.7}$   &  2.2  & 2.0 \\
$[0.6, 1.0]$                        &  1.32    &  $36^{+27}_{-26}{}^{+31}_{-26}$  &   $3.5^{+2.7}_{-2.5}{}^{+3.1}_{-2.6}$   &  9.3  & 6.4 \\
$[1.0, m_{B}]$   &  0.63    &  $25^{+45}_{-43}{}^{+63}_{-53}$  &   $5.1^{+9.2}_{-8.8}{}^{+12.9}_{-11.0}$ &  30.9 & 27.3 \\
\hline
Full range                          &  0.97    &  $66^{+64}_{-61}{}^{+95}_{-81}$  &   $8.8^{+8.5}_{-8.2}{}^{+12.6}_{-10.7}$ &  32.7 & 24.2 \\
    \hline \hline
  \end{tabular*}
\end{table}

After unveiling the signal region, which had not been examined during the optimization of the analysis, we identified and corrected an error in the single-candidate selection.
This resulted in a 2.6\% change in the mean upper limit expected from the simulation and a $0.6\%$ change in the observed upper limit from data.

In summary, we search for $B \rightarrow X_{s} \nu \bar{\nu}$ decay using $365.4\invfb$ of $\FourS$ and $42.7\invfb$ of off-resonance data collected by the Belle II experiment.  The analysis reconstructs one $B$ meson in a hadronic decay mode and considers 30 exclusive channels in the reconstruction of the $X_{s}$ system.
No significant signal is observed, and we set 90\% C.L. upper limits on the partial branching fractions of $2.2\times 10^{-5}$ for $0.0 < M_{X_{s}}^{\textrm{true}} < 0.6\gevcc$, $9.3\times 10^{-5}$ for $0.6 < M_{X_{s}}^{\textrm{true}} < 1.0\gevcc$, and $3.1\times 10^{-4}$ for $1.0\gevcc < M_{X_{s}}^{\textrm{true}}$.
We calculate the branching fraction for the entire $M_{X_{s}}^{\textrm{true}}$ region by summing the partial branching fractions to find
\begin{align*}
    {\mathcal{B}}(B \rightarrow X_{s} \nu \bar{\nu}) = [8.8^{+8.5}_{-8.2}(\textrm{stat})^{+12.6}_{-10.7}(\textrm{syst})]\times 10^{-5},
\end{align*}
and set a corresponding 90\% C.L. upper limit
\begin{align*}
    {\mathcal{B}}(B \rightarrow X_{s} \nu \bar{\nu})  < 3.3\times 10^{-4}.
\end{align*}
This branching fraction and upper limit do not include the contribution from long-distance $B^{+} \rightarrow \tau^{+}(\rightarrow X_{s}^{+}\bar{\nu})\nu$ decays.
This is the first search for inclusive $B \rightarrow X_{s} \nu \bar{\nu}$ decays
and can be used to constrain potential contributions from physics beyond the SM~\cite{Felkl:2021uxi}. 

\begin{acknowledgments}
This work, based on data collected using the Belle II detector, which was built and commissioned prior to March 2019,
was supported by
Higher Education and Science Committee of the Republic of Armenia Grant No.~23LCG-1C011;
Australian Research Council and Research Grants
No.~DP200101792, 
No.~DP210101900, 
No.~DP210102831, 
No.~DE220100462, 
No.~LE210100098, 
and
No.~LE230100085; 
Austrian Federal Ministry of Education, Science and Research,
Austrian Science Fund (FWF) Grants
DOI:~10.55776/P34529,
DOI:~10.55776/J4731,
DOI:~10.55776/J4625,
DOI:~10.55776/M3153,
and
DOI:~10.55776/PAT1836324,
and
Horizon 2020 ERC Starting Grant No.~947006 ``InterLeptons'';
Natural Sciences and Engineering Research Council of Canada, Digital Research Alliance of Canada, and Canada Foundation for Innovation;
National Key R\&D Program of China under Contract No.~2024YFA1610503,
and
No.~2024YFA1610504
National Natural Science Foundation of China and Research Grants
No.~11575017,
No.~11761141009,
No.~11705209,
No.~11975076,
No.~12135005,
No.~12150004,
No.~12161141008,
No.~12405099,
No.~12475093,
and
No.~12175041,
and Shandong Provincial Natural Science Foundation Project~ZR2022JQ02;
the Czech Science Foundation Grant No. 22-18469S,  Regional funds of EU/MEYS: OPJAK
FORTE CZ.02.01.01/00/22\_008/0004632 
and
Charles University Grant Agency project No. 246122;
European Research Council, Seventh Framework PIEF-GA-2013-622527,
Horizon 2020 ERC-Advanced Grants No.~267104 and No.~884719,
Horizon 2020 ERC-Consolidator Grant No.~819127,
Horizon 2020 Marie Sklodowska-Curie Grant Agreement No.~700525 ``NIOBE''
and
No.~101026516,
and
Horizon 2020 Marie Sklodowska-Curie RISE project JENNIFER2 Grant Agreement No.~822070 (European grants);
L'Institut National de Physique Nucl\'{e}aire et de Physique des Particules (IN2P3) du CNRS
and
L'Agence Nationale de la Recherche (ANR) under Grant No.~ANR-23-CE31-0018 (France);
BMFTR, DFG, HGF, MPG, and AvH Foundation (Germany);
Department of Atomic Energy under Project Identification No.~RTI 4002,
Department of Science and Technology,
and
UPES SEED funding programs
No.~UPES/R\&D-SEED-INFRA/17052023/01 and
No.~UPES/R\&D-SOE/20062022/06 (India);
Israel Science Foundation Grant No.~2476/17,
U.S.-Israel Binational Science Foundation Grant No.~2016113, and
Israel Ministry of Science Grant No.~3-16543;
Istituto Nazionale di Fisica Nucleare and the Research Grants BELLE2,
and
the ICSC – Centro Nazionale di Ricerca in High Performance Computing, Big Data and Quantum Computing, funded by European Union – NextGenerationEU;
Japan Society for the Promotion of Science, Grant-in-Aid for Scientific Research Grants
No.~16H03968,
No.~16H03993,
No.~16H06492,
No.~16K05323,
No.~17H01133,
No.~17H05405,
No.~18K03621,
No.~18H03710,
No.~18H05226,
No.~19H00682, 
No.~20H05850,
No.~20H05858,
No.~22H00144,
No.~22K14056,
No.~22K21347,
No.~23H05433,
No.~26220706,
and
No.~26400255,
and
the Ministry of Education, Culture, Sports, Science, and Technology (MEXT) of Japan;  
National Research Foundation (NRF) of Korea Grants
No.~2021R1-F1A-1064008, 
No.~2022R1-A2C-1003993,
No.~2022R1-A2C-1092335,
No.~RS-2016-NR017151,
No.~RS-2018-NR031074,
No.~RS-2021-NR060129,
No.~RS-2023-00208693,
No.~RS-2024-00354342
and
No.~RS-2025-02219521,
Radiation Science Research Institute,
Foreign Large-Size Research Facility Application Supporting project,
the Global Science Experimental Data Hub Center, the Korea Institute of Science and
Technology Information (K25L2M2C3 ) 
and
KREONET/GLORIAD;
Universiti Malaya RU grant, Akademi Sains Malaysia, and Ministry of Education Malaysia;
Frontiers of Science Program Contracts
No.~FOINS-296,
No.~CB-221329,
No.~CB-236394,
No.~CB-254409,
and
No.~CB-180023, and SEP-CINVESTAV Research Grant No.~237 (Mexico);
the Polish Ministry of Science and Higher Education and the National Science Center;
the Ministry of Science and Higher Education of the Russian Federation
and
the HSE University Basic Research Program, Moscow;
University of Tabuk Research Grants
No.~S-0256-1438 and No.~S-0280-1439 (Saudi Arabia), and
Researchers Supporting Project number (RSPD2025R873), King Saud University, Riyadh,
Saudi Arabia;
Slovenian Research Agency and Research Grants
No.~J1-50010
and
No.~P1-0135;
Ikerbasque, Basque Foundation for Science,
State Agency for Research of the Spanish Ministry of Science and Innovation through Grant No. PID2022-136510NB-C33, Spain,
Agencia Estatal de Investigacion, Spain
Grant No.~RYC2020-029875-I
and
Generalitat Valenciana, Spain
Grant No.~CIDEGENT/2018/020;
The Knut and Alice Wallenberg Foundation (Sweden), Contracts No.~2021.0174, No.~2021.0299, and No.~2023.0315;
National Science and Technology Council,
and
Ministry of Education (Taiwan);
Thailand Center of Excellence in Physics;
TUBITAK ULAKBIM (Turkey);
National Research Foundation of Ukraine, Project No.~2020.02/0257,
and
Ministry of Education and Science of Ukraine;
the U.S. National Science Foundation and Research Grants
No.~PHY-1913789 
and
No.~PHY-2111604, 
and the U.S. Department of Energy and Research Awards
No.~DE-AC06-76RLO1830, 
No.~DE-SC0007983, 
No.~DE-SC0009824, 
No.~DE-SC0009973, 
No.~DE-SC0010007, 
No.~DE-SC0010073, 
No.~DE-SC0010118, 
No.~DE-SC0010504, 
No.~DE-SC0011784, 
No.~DE-SC0012704, 
No.~DE-SC0019230, 
No.~DE-SC0021274, 
No.~DE-SC0021616, 
No.~DE-SC0022350, 
No.~DE-SC0023470; 
and
the Vietnam Academy of Science and Technology (VAST) under Grants
No.~NVCC.05.02/25-25
and
No.~DL0000.05/26-27.

These acknowledgements are not to be interpreted as an endorsement of any statement made
by any of our institutes, funding agencies, governments, or their representatives.

We thank the SuperKEKB team for delivering high-luminosity collisions;
the KEK cryogenics group for the efficient operation of the detector solenoid magnet and IBBelle on site;
the KEK Computer Research Center for on-site computing support; the NII for SINET6 network support;
and the raw-data centers hosted by BNL, DESY, GridKa, IN2P3, INFN, 
and the University of Victoria.

\end{acknowledgments}

\textit{Data availability} ---
Numerical data corresponding to the results presented are available as HEPData~\cite{hepdata.169553}.
The full Belle II data are not publicly available.
The collaboration will consider requests for access to the data that support this article.

\ifthenelse{\boolean{wordcount}}%
{ \nobibliography{references} }
{ \bibliography{references} }

\end{document}


\title{
First search for $B \rightarrow X_{s} \nu \bar{\nu}$ decays
}


\maketitle

\noindent \section{List of BDT input variables}
Superscripts $*$ denote variables calculated in the c.m. frames. Otherwise, they are calculated in the laboratory frame.

\subsection{Variables related to $X_{s}$ and $B_{\textrm{tag}}$ candidates} 
\begin{itemize}
  \item The magnitude of momentum $| \vec{p}_{X_{s}}^{*} |$ of the $X_{s}$ candidate
  \item The polar angle $\theta_{B_{\textrm{tag}}}^{*}$ of the $B_{\textrm{tag}}$ candidate
  \item The magnitude of the thrust vector $|\vec{T}_{X_{s}}^{*}|$ for $X_{s}$
\end{itemize}

\subsection{Variables related to rests of event~(ROE)}
We define rests of the event~(ROE) for three systems.
Each $\textrm{ROE}(S)$ with $S=X_{s}$, $B_{\textrm{tag}}$, $X_{s}+B_{\rm tag}$ contains photon candidates, KLM clusters, and tracks that are not used in the reconstruction of the $S$ system.
If a track has the highest PID likelihood for a kaon, proton, electron, or muon, it is identified as that particle; otherwise, it is identified as a pion.
Tracks included in the ROEs must satisfy $dr < 0.5\cm$ and $\left| dz \right| < 2\cm$, and to be within the CDC acceptance region.
Additionally, tracks must have more than 20 hits in the CDC.
The energy of photon candidates in the ROE must be greater than 100, 60, and 150\mev in the forward, barrel, and backward regions of the ECL, respectively.
\begin{itemize}
  \item $\theta$ of $\sum{\vec{p}}$ of photon candidates used for $E_{\textrm{extra}}$
  \item The cosine $\hat{T}^{*}_{X_{s}} \cdot \hat{T}_{\textrm{ROE}(X_{s})}^{*}$ of the angle between the thrust axes of the $X_{s}$ candidate and the $\textrm{ROE}(X_{s})$
  \item $\hat{T}_{B_{\textrm{tag}}}^{*} \cdot \hat{T}_{\textrm{ROE}(B_{\textrm{tag}})}^{*}$
  \item Modified Fox-Wolfram moments~\cite{Belle:2003fgr} in the c.m. frame for particles in $B_{\textrm{tag}}$ and $\textrm{ROE}(B_{\textrm{tag}})$
  \item Modified Fox-Wolfram moments~\cite{Belle:2003fgr} in the c.m. frame for particles in $X_{s}$ and $\textrm{ROE}(X_{s})$
  \item The first, second, and third CLEO cones in the c.m. frame~\cite{CLEO:1995rok} for particles in $B_{\textrm{tag}}$ and $\textrm{ROE}(B_{\textrm{tag}})$
  \item The number of photon candidates in the $\textrm{ROE}(X_{s}+B_{\textrm{tag}})$
  \item $E_{\textrm{extra}}$
\end{itemize}

\subsection{Variables related to the event shape}
Event shape variables are obtained using all tracks and photons.
The tracks must satisfy $|dz| < 3~\textrm{cm}$, $dr < 0.5~\textrm{cm}$, and be within the CDC acceptance region.
Their transverse momentum must be greater than $0.1\gevc$.
The photons must also be within the CDC acceptance region.
The ECL cluster time and the event time must not differ by more than $200\ns$.
The photon energy must be greater than $50\mev$.
\begin{itemize}
  \item $| \vec{p}_{\textrm{miss}} |$
  \item $\theta_{\textrm{miss}}$
  \item The third and fourth normalized Fox-Wolfram moments in the c.m. frame~\cite{Fox:1978vw}
  \item The first and second-order harmonic moments with respect to the thrust axis in the c.m. frame~\cite{Fox:1978vw}
\end{itemize}

\subsection{Variables related to background from $D$ mesons}
To suppress background from $D$ mesons, we form $D$ meson candidates from a kaon from the $X_{s}$ candidate and pions from the entire event, in three categories.
The first category is $D^+$ decays without a $\pi^{0}$ :
$D^{+}\rightarrow K_{S}^{0} \pi^{+}$, $D^{+}\rightarrow K^{-} \pi^{+} \pi^{+}$.
The second category is $D^0$ decays without a $\pi^0$ :
$D^{0} \rightarrow K^{-} \pi^{+}$, $D^{0} \rightarrow K_{S}^{0} \pi^{+} \pi^{-}$, and $D^{0} \rightarrow K^{-} \pi^{+} \pi^{-} \pi^{+}$.
The third category is $D^0$ decay modes with or without $\pi^{0}$:
$D^{0} \rightarrow K^{-} \pi^{+}$, $D^{0} \rightarrow K_{S}^{0} \pi^{+} \pi^{-}$, $D^{0} \rightarrow K^{-} \pi^{+} \pi^{0}$, $D^{0} \rightarrow K_{S}^{0} \pi^{+} \pi^{-} \pi^{0}$, and $D^{0} \rightarrow K^{-} \pi^{+} \pi^{0} \pi^{0}$.
After the reconstruction, a vertex fit is performed and is required to converge.
We use the following six variables as $D$-veto inputs for the BDT. If no $D$ meson candidate is found, a value of 0 is assigned:
\begin{itemize}
  \item The standard deviation of the vertex-fit $p$-values among $D$ meson candidates for each category
  \item The mass of the $D$-meson candidate that has the highest vertex-fit $p$-value in each category
\end{itemize}

\subsection{Other variable}
\begin{itemize}
  \item The number of tracks not consistent with originating from the IP and not used for the $X_{s}$ or $B_{\rm tag}$ candidates
\end{itemize}

\section{Corrections and Minor Systematic Uncertainty}
  
\begin{table*}[]
\renewcommand{\arraystretch}{1.3}
\caption{Summary of systematic uncertainties on the ${\cal{B}}(B \to X_s \nu \bar{\nu})$ in each $M_{X_{s}}^{\textrm{true}}$ range. The impact on the branching fraction uncertainty $\sigma_{\cal{B}}$ from each source is estimated by fixing the corresponding nuisance parameter in the minimization procedure and subtracting its uncertainty in quadrature from the total uncertainty. Due to correlations among sources, the quadrature sum of individual systematic uncertainties does not equal the total systematic uncertainty. $M_{X_{s}}^{\textrm{max}}$ is the mass of the $B$ meson.}
\label{tab:uncertaintiestable}
\centering
\begin{tabular*}{\textwidth}{@{\extracolsep{\fill}}lccccc}
 \hline
  \hline
       & & \multicolumn{4}{c}{Impact on $\sigma_{\cal{B}}$ [$10^{-5}$]} \\\cline{3-6}
Source & Uncertainty size & Full range & $[0, 0.6]\gevcc$ & $[0.6, 1.0]\gevcc$ & $[1.0, M_{X_{s}}^{\textrm{max}}]\gevcc$ \\
 \hline
Simulated-sample size & $O(1\%)$ & 6.0 & 0.4 & 1.5 & 6.3 \\
Background normalization & 20\% & 5.7 & 0.6 & 2.0 & 6.6 \\
Branching fraction of major $B$-decay & $O(1 \text{-} 100\%)$ & 2.3 & 0.1 & 0.6 & 2.0 \\
Non-resonant $X_{s}\nu\bar{\nu}$ generation range & 9.1\% & 2.1 & 0.0 & 0.0 & 2.2 \\
$\mathcal{O}_{\textrm{BDT}}$ selection efficiency & $O(1 \text{-} 10\%)$ & 2.1 & 0.0 & 0.4 & 2.3 \\
Photon multiplicity correction & $O(1\%)$ & 1.8 & 0.0 & 0.2 & 1.6 \\
$q\bar{q}$ background efficiency & $O(10\%)$ & 1.8 & 0.1 & 0.2 & 1.8 \\
Fitting bias & $O(5\%)$ & 1.6 & 0.2 & 0.1 & 1.7 \\
Tagging efficiency & $O(1 \text{-} 30\%)$ & 1.1 & 0.1 & 0.3 & 0.8 \\
Tagging efficiency from $B \to K^{(*)}J/\psi$ decay & $O(20 \text{-} 100\%)$ & 1.0 & 0.1 & 0.4 & 1.0 \\
${\mathcal{O}_{\textrm{BDT}}}$ distribution shape for $q\bar{q}$ background & 100\% of weight & 1.0 & 0.1 & 0.2 & 0.7 \\
Branching fraction for $B \to X_{s} K_{L}^{0} K_{L}^{0}$ & 100\% & 0.9 & 0.0 & 0.3 & 0.6 \\
The number of $B\Bbar$ pair & 1.45\% & 0.7 & 0.0 & 0.1 & 0.6 \\
Fermi motion momentum & $O(8\%)$ & 0.7 & 0.0 & 0.0 & 0.7 \\
Fragmentation & $O(1 \text{-} 100\%)$ & 0.6 & 0.0 & 0.1 & 0.7 \\
$B\to K^{*}$ form factor & $O(10\%)$ & 0.6 & 0.0 & 0.5 & 0.6 \\
Branching fraction for $B \to X_{s} n\bar{n}$ & 100\% & 0.5 & 0.1 & 0.4 & 0.2 \\
Fraction of $K\nu\bar{\nu}$, $K^{*}\nu\bar{\nu}$ decays & $O(5\%)$ & 0.4 & 0.1 & 0.3 & 0.4 \\
b-quark mass parameter & 3.2\% & 0.3 & 0.0 & 0.0 & 0.3 \\
Branching fraction for $D \to X K_{L}^{0}$ from $B$ meson & 7.7\% & 0.3 & 0.0 & 0.1 & 0.2 \\
Charged pion identification & $O(1\%)$ & 0.2 & 0.0 & 0.1 & 0.2 \\
$B\to K$ form factor & $O(1\%)$ & 0.2 & 0.1 & 0.0 & 0.2 \\
Tracking efficiency & 0.27\% & 0.2 & 0.0 & 0.0 & 0.1 \\
$\pi^{0}$ reconstruction efficiency & $O(1\%)$ & 0.1 & 0.0 & 0.0 & 0.1 \\
Charged kaon identification & $O(1\%)$ & 0.1 & 0.0 & 0.0 & 0.1 \\
$K_{S}^{0}$ reconstruction efficiency & $O(5\%)$ & 0.1 & 0.0 & 0.0 & 0.1 \\
$K_{L}^{0}$ efficiency in the ECL & 17\% & 0.1 & 0.0 & 0.0 & 0.1 \\
\hline
Total systematic sources & & 11.7 & 0.8 & 2.8 & 12.0 \\
\hline
Statistical uncertainty & & 8.3 & 0.8 & 2.6 & 9.0 \\

 \hline
 \hline
\end{tabular*}
\end{table*}

Table~\ref{tab:uncertaintiestable} summarizes the systematic uncertainties in each $M_{X_{s}}^{\textrm{true}}$ region. In the main text, the following sources of systematic uncertainties are discussed: simulated-sample size, background normalization, branching fraction of major $B$-decay, non-resonant $X_{s}\nu\bar{\nu}$ generation point, $\mathcal{O}_{\textrm{BDT}}$ selection efficiency, photon multiplicity correction, $q\bar{q}$ background efficiency, and ${\mathcal{O}_{\textrm{BDT}}}$ distribution shape for $q\bar{q}$ background. The remaining systematic sources are discussed here.

The systematic uncertainty from fit bias is estimated using 10000 simulated samples.
Based on the mean shifts between the fitted and input values, branching fraction uncertainties of $18.4\%$, $3.5\%$, and $9.4\%$ are assigned for the mass regions $M_{X_s}^{\textrm{reco}} < 0.6\gevcc$, $0.6 < M_{X_s}^{\textrm{reco}} < 1.0\gevcc$, and $1.0 < M_{X_s}^{\textrm{reco}} < 2.0\gevcc$, respectively.
Additionally, based on the discrepancy between the fit errors and the observed spread of fitted values, uncertainties of $1.1\%$, $0.9\%$, and $2.9\%$ are assigned for the same mass regions.

To correct discrepancies in the FEI algorithm between data and simulation, $B \rightarrow X \ell \nu$ decays are studied~\cite{Belle-II:2020fst}.
Correction factors and their associated uncertainties are applied as functions of $B_{\textrm{tag}}$ decay modes.

The FEI correction factor is estimated from $B \rightarrow X \ell \nu$ decays.
Since the signal-side $B$ meson decays differ, the $B_{\textrm{tag}}$ efficiency may vary.
To account for this, we compare the data-to-simulation efficiency ratio using $B \rightarrow K^{*} J/\psi$ events.
The difference between the FEI calibration factors obtained from $B \rightarrow X \ell \nu$ and $B \rightarrow K^{*} J/\psi$ is assigned as an additional systematic uncertainty.
The statistical uncertainty from the $B \rightarrow K^{*} J/\psi$ sample is also included as a systematic uncertainty.

A systematic uncertainty of $\pm100\%$ is assigned to the branching fractions of the $B \rightarrow X_{s} K_{L}^{0} K_{L}^{0}$ decays to cover mismodeling.

The simulated samples are scaled using $N_{\Upsilon(4S)} = (387.1 \pm 5.6) \times 10^{6}$, and a 1.5\% relative uncertainty is assigned from the number of $B\Bbar$ pairs.

In the Fermi motion model, the Fermi momentum $p_{F} = 0.461^{+0.036}_{-0.039}\gevc$\cite{Kim:1998wx,HFLAV:2022pwe} is used as a parameter.
We vary $p_{F}$ within its uncertainty, and this variation is taken as the systematic uncertainty.

The systematic uncertainty of the fragmentation is estimated from the differences between fragmentation in $B \rightarrow X_{s} \gamma$~\cite{PhysRevD.99.032012} and $B \rightarrow X_{s} J/\psi$~\cite{Belle:2014owz} decays.
These uncertainties are evaluated in two $M_{X_{s}}^{\textrm{true}}$ regions, $[1.15, 1.5]\gevcc$, $[1.5, 2.0]\gevcc$, and for seven decay mode categories: $K\pi$ without $\pi^{0}$, $K\pi$ with one $\pi^{0}$, $K2\pi$ without $\pi^{0}$, $K2\pi$ with one $\pi^{0}$, $K3\pi$ without $\pi^{0}$, $K3\pi$ with one $\pi^{0}$, and $K4\pi$ at most one $\pi^{0}$.
To estimate the systematic uncertainty for the remaining decay modes, Pythia parameters are varied such that the simulated fragmentation in the seven categories remains within twice the observed differences between the $B \rightarrow X_{s} \gamma$ and $B \rightarrow X_{s} J/\psi$ results.
The maximum deviation under these variations is taken as the systematic uncertainty for the remaining reconstructed decay modes and for unreconstructed modes.

The uncertainties of the SM $B \rightarrow K^{*}$ form factors are taken from Refs.~\cite{Bharucha:2015bzk}, and are used to estimate the systematic uncertainties associated with the form factors in the $B \rightarrow K^{*} \nu \bar{\nu}$ decay.

A systematic uncertainty of $\pm100\%$ is assigned to the branching fractions of the $B \rightarrow X_{s} n\bar{n}$ decays to cover mismodeling.

There is an uncertainty associated with the relative fractions of $B \rightarrow K \nu \bar{\nu}$ and $B \rightarrow K^{*} \nu \bar{\nu}$ within the total $B \rightarrow X_{s} \nu \bar{\nu}$ decay.
Since the uncertainties from the form factors are treated separately, we evaluate the branching fraction uncertainties from other parameters.
Relative uncertainties of $5.7\%$ and $5.4\%$ are assigned to the branching fractions of $B \rightarrow K \nu \bar{\nu}$ and $B \rightarrow K^{*} \nu \bar{\nu}$, respectively~\cite{Parrott:2022rgu,Bharucha:2015bzk}.

The $b$-quark mass is used in the calculation of the $q^{2}$ distribution for the $B \rightarrow X_{s} \nu \bar{\nu}$ decay.
We use $m_{b} = 4.68 \pm 0.15\gevcc$, which covers the central values used in the theoretical calculations of $B \rightarrow X_{s} \nu \bar{\nu}$~\cite{Altmannshofer:2009ma} and $B \rightarrow K^{(*)} \ell \ell$~\cite{Ali:1999mm} decays. We vary $m_{b}$ within its uncertainty, and the variation of $q^{2}$ distribution is applied.

According to the study of $B^{+} \rightarrow K^{+} \nu \bar{\nu}$ decay, applying a correction factor of $1.3 \pm 0.1$ to decays involving the $B \rightarrow D \rightarrow K_{L}^{0}$ transition reduces the discrepancy between data and simulations~\cite{Belle-II:2023esi}.
We adopt this correction factor and its associated uncertainty in our analysis.

The efficiency and misidentification rates for pions are obtained using the $D^{*+} \rightarrow D^{0}(\rightarrow K^{-} \pi^{+}) \pi^{+}$ decay.
Correction factors and their associated uncertainties are applied.

The uncertainties of the SM $B \rightarrow K$ form factors are taken from Refs.~\cite{Parrott:2022rgu}, and are used to estimate the systematic uncertainties associated with the form factors in the $B \rightarrow K \nu \bar{\nu}$ decay.

The systematic uncertainty from the tracking efficiency is estimated using $e^{+}e^{-} \rightarrow \tau^{+}\tau^{-}$ events, in which one $\tau$ lepton decays into three charged tracks and the other decays into a single charged track.

To correct the $\pi^{0}$ reconstruction efficiency, the $D^{0} \rightarrow K^{-} \pi^{+} \pi^{0}$ and $D^{0} \rightarrow K^{-} \pi^{+}$ decays are studied.
Correction factors and their associated uncertainties are applied as functions of $\pi^{0}$ momentum.

The efficiency and misidentification rates for kaons are obtained using the $D^{*+} \rightarrow D^{0}(\rightarrow K^{-} \pi^{+}) \pi^{+}$ decay.
Correction factors and their associated uncertainties are applied.

The systematic uncertainty associated with the $K_{S}^{0}$ reconstruction efficiency is estimated using $D^{*+} \rightarrow D^{0}(\rightarrow K_{S}^{0} \pi^{+}\pi^{-}) \pi^{+}$ decays.

From the study of the $B^{+} \rightarrow K^{+} \nu \bar{\nu}$ decay, a systematic uncertainty on the efficiency of ECL clusters from $K_{L}^{0}$ is required to account for the observed discrepancy between data and simulations~\cite{Belle-II:2023esi}.
We adopt this systematic uncertainty and assign a $\pm17\%$ uncertainty to the efficiency of ECL clusters from $K_{L}^{0}$.
\section{SUPPLEMENTAL FIGURES}
\begin{figure}[h]
    \includegraphics[width=\columnwidth]{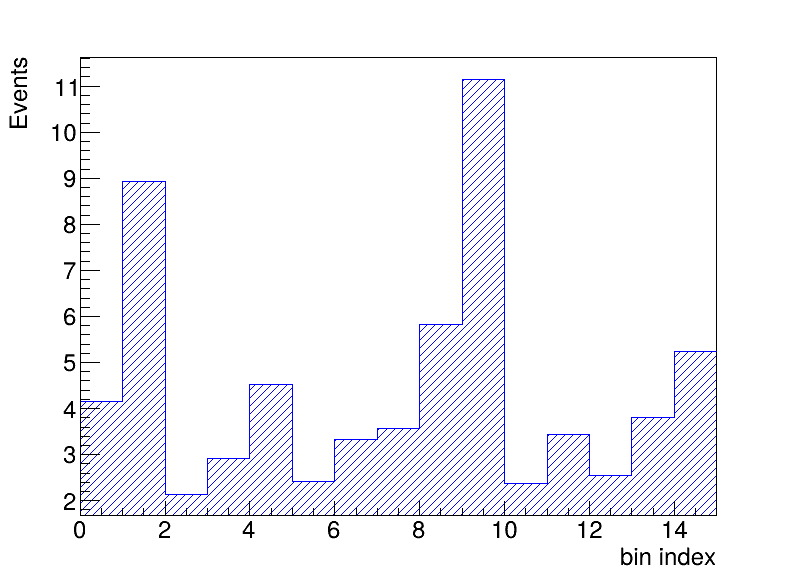}
    \caption{The bin index distribution after the fit, for the signal template.}
\end{figure}
\begin{figure}[h]
    \includegraphics[width=\columnwidth]{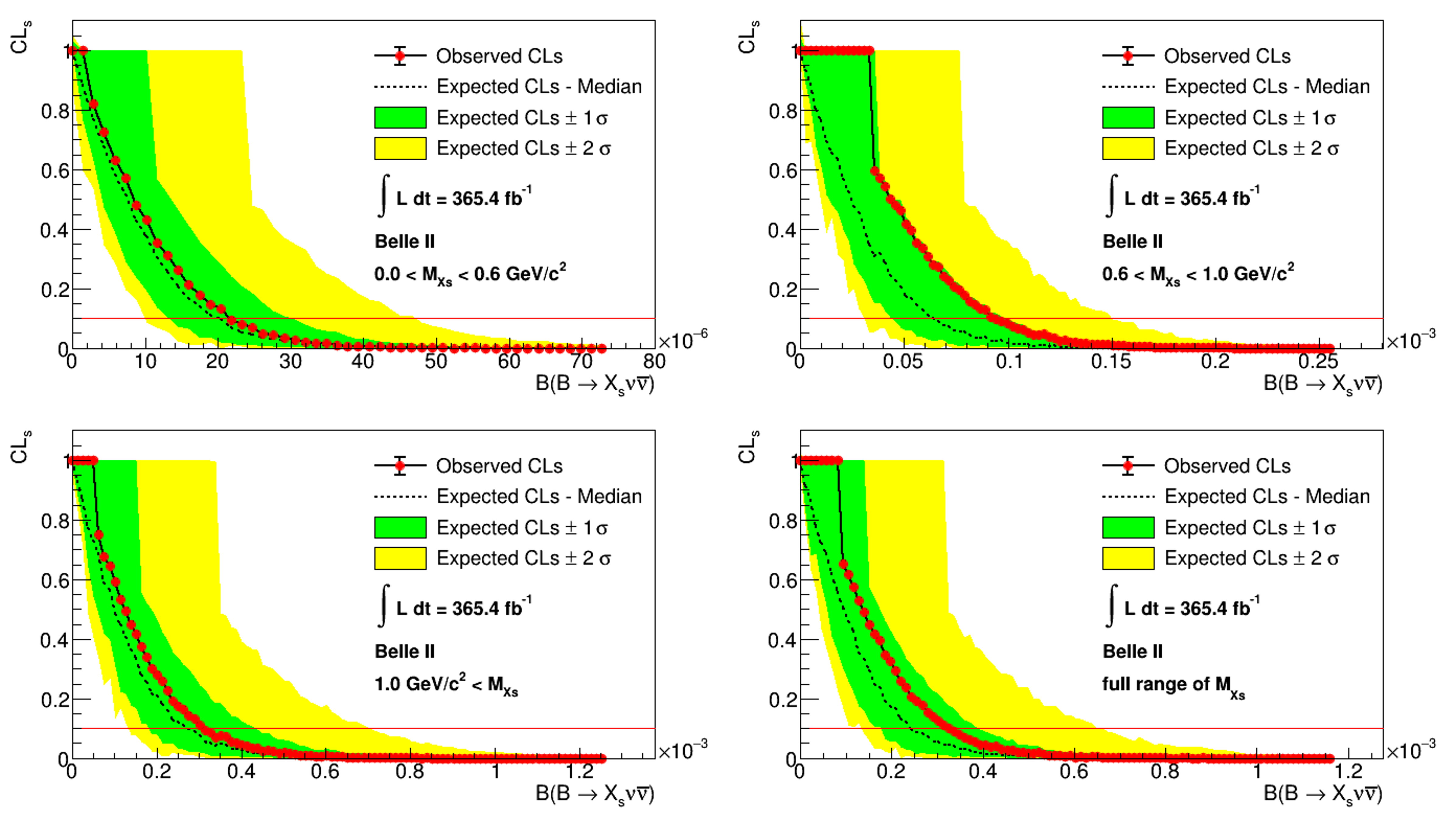}
    \caption{$\textrm{CL}{s}$ values as a function of the assumed branching fraction for $B \rightarrow X{s} \nu \bar{\nu}$ in each $M_{X_{s}}^{\textrm{true}}$ region. The red line indicates the 90\% confidence level upper limits.}
\end{figure}
\ifthenelse{\boolean{wordcount}}%
{ \nobibliography{references} }
{ \bibliography{references} }